\documentstyle[12pt,epsf,psfig]{l-aa}     

\begin{document}
\thesaurus{ 03  (11.01.2; 11.05.2; 11.19.7; 13.09.1.) }

\title{ Deep Galaxy survey at 6.75$\mu$m with the ISO satellite.
  \thanks{Based on observations with ISO, an ESA project with instruments
    funded by ESA Member States (especially the  countries: France,
    Germany, the Netherlands and the United Kingdom) with the
    participation of ISAS and NASA. }
  }

   \author{H. Flores\inst{1}, F. Hammer\inst{1}, F.X. D\'esert\inst{2},
     C. C\'esarsky\inst{3}, T. Thuan\inst{4}, D. Crampton\inst{5},
     S. Eales\inst{6},  O. Le F\`evre\inst{1},
     S.J. Lilly\inst{7}, A. Omont\inst{8} and D. Elbaz\inst{3},}
   \offprints{H. Flores; Observatoire de Paris, F-92195 Meudon Cedex,
     France}
   \institute{Observatoire de Meudon, DAEC and URA 173,
     associ\'e au CNRS
     et \`a l'Universit\'e Paris 7, 92195 Meudon Cedex, France
     \and
     Institut d'Astrophysique Spatiale, Orsay. France.
     \and
     Service d'Astrophysique, CEA, France.
     \and
     University of Virginia, USA.
     \and
     Dominion Astrophysical Observatory, HIA, National Research Council
     of Canada, V8X 4M6, Canada
     \and
     University of Cardiff, UK.
     \and
     Department of Astronomy, University of Toronto, Toronto, Canada.
     \and
     Institute d'Astrophysique de Paris, France.}
   \date{Received : 5 January 1998; accepted : 02 September 1998}
   \markboth{H. Flores et al.}{ISOCAM LW2 catalogue. }
\maketitle

\begin{abstract}
Deep 6.75$\mu$m mid-IR ISOCAM observations were obtained of the
Canada-France Redshift Survey (CFRS) 1415+52 field with the Infrared
Space Observatory. The identification of the sources with optical
counterparts is described in detail, and a classification scheme is
devised which depends on the S/N of the detection and the inverse
probability of chance coincidence.  83\% of the 54 ISOCAM sources are
identified  with $I_{AB} <$23.5 counterparts.  The  $(I-K)_{AB}$
colors, radio properties,  spectrophotometric properties and frequency
of nuclear activity  of these counterparts differ on average from those
of typical CFRS galaxies.  CFRS spectra are available for 21 of the
sources which have I$_{AB} \leq 22.5$ (including 7 stars).  Most of the
strongest sources are stars or AGN. Among the non--stellar counterparts
with spectra, 40\% are AGNs, and 53\% are galaxies that display star
formation activity and/or significant contributions of A stars.
The ISOCAM sources also display an IR excess, even when compared with
heavily-reddened local starburst galaxies. An upper limit of 30\% of
extragalactic ISO sources could be at $ z > 1$ of the 44
$S_{6.75\mu m} > 150\mu Jy$ sources which are non-stellar
(7 "spectroscopic" and 3 "photometric" stars excluded).
\end{abstract}

\section{Introduction}
Very little is known about the mid-IR (5 $\leq \lambda \leq $ 40$\mu$m)
properties of distant galaxies, wavelengths which should probe their
dust content and star formation activity.  IRAS data established that
mid-IR emission from the interstellar medium and from nearby star
forming galaxies is dominated by emission from Very Small grains and
polycyclic aromatic hydrocarbon molecules (PAHs) which have
fluctuating temperatures under single photon absorption, whereas classical
grains are in thermal equilibrium and emit at longer wavelengths (Helou 1986).
However, the precise nature of PAHs remains open (Puget \& L\'eger
1989).
Differences in mid-IR features among various galaxies can be
attributed to differing amounts of dust, broad emission interpreted
as due to PAHs or Unidentified Infrared Bands (UIBs) carriers.
The PAH and very small grains appear to be responsible for
$\sim$30\% of the total IR emission  in normal galaxies, but in active
galactic nuclei, the mid-IR emission is believed to arise from  a dusty torus
around the active nucleus where PAHs are destroyed (Edelson \& Malkan
1986, Roche et al. 1991, Helou et al., 1991).

In order to study the mid-IR emission from high redshift galaxies, deep
ISO (Infrared Space Observatory, Kessler et al. 1996) observations were
made of the CFRS field at 1415+52.  Extensive spectroscopic and $B, V,
I$ and $K$ photometric data already exist (Lilly et al. 1995a,b) for
galaxies in this field, as well as data from a deep (S$_{5GHz} \geq$ 16
$\mu$Jy) radio survey by Fomalont et al. (1991). The high spatial
resolution  in the micro-scanning mode, combined with the good
sensitivity of CAM (C\'esarsky et al. 1996) allows mid-IR maps of high
redshift field galaxies to be made for the first time. Even so, precise
identification of faint ISO sources with such galaxies is difficult,
owing to their faintness and to the inherent uncertainties in the
source positions.

The layout of this  paper is as follows; Sect. 2 presents the
observational  and data reduction strategy; Sect. 3 gives a description
of the catalogue, astrometry and identification of optical
counterparts. In Sect. 4 the photometric and spectrophotometric
properties of the ISO objects are discussed. Section 5 summarizes the
results of the identification procedure and, finally, our conclusions
about the faint mid-IR sources are discussed in Sect. 6.

The nature of 6.75$\mu$m sources and their energy distribution from UV to radio wavelengths will be fully discussed in a forthcoming paper (Flores et al, 1998), which will also present the 15$\mu$m data. 

\section{Observations and  data reduction}
A $13\arcmin \times 13\arcmin$ region centered on the CFRS 1415+52
field  was observed with ISO in a raster pattern such that a constant
exposure time per pixel was achieved over the whole 10\arcmin\ $\times$
10\arcmin\ CFRS field.  Given the small size of this field, it was
decided to go as deep as possible with ISOCAM at 6.75$\mu$m, partly so
that the results could be used as a possible template for
the ISOCAM Central Program  for deep surveys (C\'esarsky et al. 1996),
as well as for other large surveys.

Eleven individual images were taken in the  micro scanning AOT mode
(CAM01, a raster of 4$\times$4 with  8 or 12 readouts per step ) with the ISOCAM
 LW channel (6\arcsec\ per pixel) and the LW2
filter ( 5--8.5$\mu$m),  leading to a total integration time of $\sim$
600 sec pixel$^{-1}$.
The micro scanning mode provides the best
spatial resolution through superposition of images.  The same pixel of
the sky was placed in different parts of the camera in order to
minimize and detect any systematic effects. The micro scanning AOT
technique also allows an accurate flat-field image to be generated and
yields a pixel size of 1\farcs5 in the final integrated image.  The
detection and removal of transients and glitches, integration of
images, absolute flux calibration, and source detection were carried
out using the method described by D\'esert et al. (1998).  This method
has been found to be
particularly well adapted to our observational strategy i.e. coadding
the eleven images, without redundancy within each image.  Special
attention was paid to possible error propagation in the flux values.
Finally, to further detect and eliminate relatively weak (S/N~$\sim$~3)
spurious sources, the data were also reduced with the CIA  PRETI
software (Aussel et al. 1997 and Starck et al. 1998)
and compared with the initial results. However latter software is less adapted to our data since we have generated several images the micro scanning technique, and glitch removing is more difficult. It is beyond the scope of this paper to compare the two data reduction procedures in order to estimate the photometric accuracy, and this question is addressed elsewhere (Desert et al., 1998). On the other hand, sources which have not recovered with PRETI software will not be further considered (sect. 2.1)  

Individual images were carefully registered with each other in order to
optimize the image quality of the brightest compact objects (mostly
stars).  However, this registration is limited by the presence of
glitches (defaults and cosmic rays).  In order to obtain the best
possible accuracy, an iterative procedure was adopted in which shifts
were determined from the average of the 3--5 highest S/N sources in
each of the 11 individual images and the composite image, the frames
were then offset and combined to form a new composite image, and the
procedure repeated three times. As demonstrated in Fig. 1, this results
in a significant improvement of the shapes and FWHM of the sources. The
final image (Fig. 2) of the whole ISO field has a resolution equivalent
to a median FWHM$\sim$11\arcsec\ (calculated with DAOPHOT under IRAF).

The precise location of the ISOCAM image relative to deep CFRS $B,V,$
and $I$ images of the 1415+52 field was determined from the six
brightest ISOCAM LW2 sources (5 stars and the z = 0.216 galaxy
CFRS14.1157).

\begin{figure}
  \begin{center}
    \leavevmode
    \psfig{file=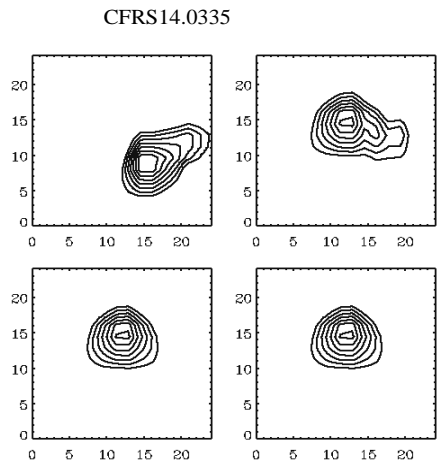,height=6cm,width=6cm}
  \end{center}
    \caption{Charts of 24''x24'' showing the changes in the contours at
      3 $\sigma$ of an ISOCAM LW2 source
       (a bright star) resulting from the iterative registration procedure.
        The first iteration is at top left, the last is at bottom right.}
\end{figure}

\begin{figure*}
  \begin{center}
    \leavevmode
    \psfig{file=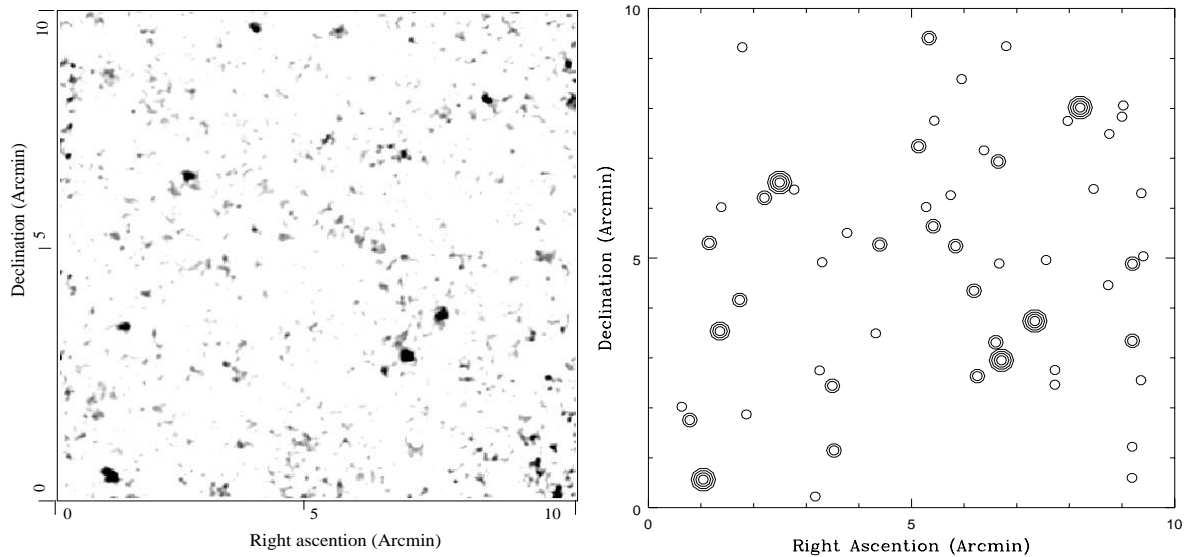,height=8cm,width=16cm}
  \end{center}
  \caption{{\it left}. The left side shows the final combined image obtained at 6.75$\mu$m with ISO of the whole 10'x10' CFRS 1415+52 field with a scale of 1.5\farcs per pixel. The image has a median resolution of FWHM$\sim$ 11\farcs. {\it righ} This diagram shows the distributions of the sources which have successfully passed our test procedure (see text). The sources with S/N $\geq$ 3 are represented by a single circle.  Sources with S/N $\geq$ 4, 6 and 8 are represented by 2, 3 and 4 concentric  open circles, respectively. Comparison of both frames provides an estimation of the noise along the map. }
\end{figure*}

Figure 3 shows the 5--8.5 micron flux distribution corrected for aperture effects ($\sim$1.4, for details, see D\'esert et al. 1998 and
references therein).At the S/N $\geq$ 4 limit, our number counts (21 per 100 arcmin$^{2}$) are
comparable to those of the Deep ISOCAM Survey ($\sim 0.22$ counts per
arcmin$^{2}$ for sources with a flux $>$ 250 $\mu$Jy, C\'esarsky et al. 1998, in prep.).
The validity of detection at S/N $\geq$ 4 is confirmed by studies
in Lockman-Hole
Deep Survey(C\'esarsky et al. 1998, in prep), which show that  sources
with S/N $\geq 4$ in individual frames are
confirmed in  95\% of cases, in the final integrated image (S/N $>$ 10)

\subsection{ISO source catalogue}

Classifications and verification of the source detections were made on
the basis of S/N and the repeatability of the detections in three
independent combinations of the 11 individual images (for details about
the source detection, repeatability  and classification, see D\'esert et
al. 1998). The repeatability test is based on the redundancy factor,
which is the number of times that the sky pixel was seen by different
pixels on the camera. The software built three independent projection
subrasters, and for each source candidate the flux and error are measured at the same position in each subrasters.
The quality factor is based on flux measurements and varies from the best confidence index (=4) to the worse confidence level (=0, see eqs. 7 to 11 of D\'esert et al.,1998). We have considered only sources with
confidence level higher than 3, which means that the final source flux and source fluxes in subrasters are within 3 $\sigma$  (2 $\sigma$ in the case  of 4), where $\sigma$ is the error value of the final flux source.\\
The final catalogued sources were also all detected in the
image which was reduced by the second independent analysis (using
PRETI). Altogether, 54 sources with S/N~$\geq$~3 met these criteria,
but only the subgroup of 23 sources with S/N~$\geq$~4 are considered to
be secure detections.  The 54 sources are listed in Tables 1 and 2. The
first two sections of Table 1 contain 21 ISO sources (catalogues 1 \&
2) that have secure optical identifications. The 24 sources in the
lower half of Table 1 and the 9 sources in Table 2 have less secure or no optical identifications, respectively.

\begin{table*}
\caption{Optical counterparts of ISOCAM LW2 sources.}
\label{obsgal}
\leavevmode
{\scriptsize
\begin{center}
\begin{tabular}{lcccccrrrr} \\ \hline\hline
ISO   & CFRS    &  z$^1$     &I$_{AB}$ &V$_{AB}$ & K$_{AB}^2$& d$^3$     & P$^4$
 & Flux$^5$  & Error\\\hline
\hline \multicolumn{10}{|c|}{Catalogue 1: Objects with S/N $>$ 4 \& 0.02 $>$ $P(
d,I)$ } \\\hline
 001 & 14.1157 &  0.220 &  20.54 &  22.45 &    --  &  1.97 & $1.x10^{-5}$ $^*$ &
  613. &   44.\\
 004 & 14.0098 &  star  &  16.44 &  14.66 &    --  &  2.08 & 0.0000583 &  601. &
   57.\\
 007 & 14.1400 &  star  &  15.61 &  16.13 &    --  &  3.73 & 0.0004195 &  329. &
   35.\\
 017 & 14.0138 &  star  &  15.77 &  16.80 &  15.43 &  4.08 & 0.0005019 &  267. &
   35.\\
 023 & 14.0335 &  star  &  14.77 &  17.98 &    --  &  1.45 & 0.0000284 &  299. &
   37.\\
 031 & 14.1265 &  star  &  15.60 &  17.63 &  14.87 &  1.41 & 0.0000600 &  312. &
   32.\\
 035 & 14.9025 &  0.155 &  18.38 &  19.15 &    --  &  1.45 & $2.x10^{-5}$ $^*$ &
  173. &   33.\\
 108 & 14.1134 &  star  &  16.45 &  18.47 &    --  &  1.53 & 0.0001570 &  199. &
   36.\\
 163 & 14.0198 &  1.603 &  20.04 &  20.21 &  19.86 &  1.79 & 0.0051602 &  156. &
   33.\\
 209 & 14.0685 &  star  &  17.19 &  17.93 &  17.23 &  4.58 & 0.0031149 &  155. &
   31.\\
 223 & 14.0836 &   ---  &  15.91 &  17.17 &    --  &  3.96 & 0.0004729 &  164. &
   37.\\
 315 & 14.9154 &  0.812 &  21.57 &  23.06 &    --  &  3.51 & $4.x10^{-4}$ $^*$
&  164. &   33.\\
 366 & 14.0051 &   ---  &  17.55 &  18.76 &    --  &  1.57 & 0.0003665 &  139. &
   32.\\
\hline \multicolumn{10}{|c|}{Catalogue 2: Objects with S/N $>$ 4 \&  0.317$>$ $P
(d,I)$ $>$ 0.02 } \\\hline
 114 & 14.0112 &   ---  &  23.53 &  24.86 &  20.61 &  2.96 & 0.1421836 &  151. &
   34.\\
 130 & 14.1565 &   ---  &  20.98 &  23.29 &  19.38 &  6.91 & 0.0741998 &  174. &
   38.\\
 136 & 14.1042 &  0.875 &  21.49 &  23.38 &  19.79 &  4.93 & 0.0831932 &  126. &
   31.\\
 242 & 14.9907 &   ---  &  22.95 &  19.41 &    --  &  4.77 & 0.1646952 &  192. &
   36.\\
 276 & 14.0972 &  0.674 &  21.17 &  21.99 &  20.34 &  3.05 & 0.0326979 &  137. &
   32.\\
 297 & 14.1567 &  0.479 &  19.79 &  20.04 &  18.62 &  5.46 & 0.0215114 &  189. &
   36.\\
 465 & 14.1129 &   ---  &  21.05 &  22.35 &  20.30 & 10.11 & 0.3059948 &  148. &
   34.\\
 477 & 14.0287 &   ---  &  22.29 &  24.19 &  19.98 &  2.77 & 0.0588822 &  164. &
   32.\\
\hline \multicolumn{10}{|c|}{Catalogue 4: Objects with 4 $>$S/N $>$ 3 \& $P(d,I)
$ $>$ 0.02} \\\hline
 097 & 14.1139 &  0.660 &  20.20 &  21.49 &  18.92 &  0.51 & $6.x10^{-5}$ $^*$ &
  115. &   34.\\
 178 & 14.1329 &  0.375 &  20.60 &  19.52 &    --  &  9.42 & $2.x10^{-5}$ $^*$ &
  109. &   33.\\
 187 & 14.0573 &  0.010 &  16.90 &  17.09 &  17.53 &  3.33 & $7.x10^{-4}$ $^*$ &
  124. &   32.\\
 188 & 14.0667 &   ---  &  20.17 &  19.48 &  18.92 &  9.09 & $1.x10^{-4}$ $^*$ &
  131. &   40.\\
 196 & 14.1080 &  0.066 &  20.34 &  20.68 &    --  &  1.38 & 0.0030702 &  112. &
   37.\\
 403 & 14.1303 &  0.985 &  19.97 &  19.88 &  19.25 &  2.25 & $2.x10^{-4}$ $^*$ &
  113. &   34.\\
 461 & 14.0861 &   ---  &  23.02 &  23.52 &  20.23 &  9.51 & 0.0000640 &  107. &
   35.\\
\hline \multicolumn{10}{|c|}{Catalogue 5: Objects with 4 $>$S/N $>$ 3 \& 0.317$>
$ $P(d,I)$ $>$ 0.02} \\\hline
 032 & 14.1598 &   ---  &  19.84 &  21.00 &    --  &  6.32 & 0.0287157 &  173. &
   47.\\
 055 & 14.1561 &   ---  &  19.55 &  20.06 &    --  & 10.83 & 0.0819987 &  213. &
   59.\\
 064 & 14.1563 &   ---  &  22.78 &  23.56 &    --  &  5.84 & 0.2364296 &  201. &
   55.\\
 189 & 14.0896 &   ---  &  22.40 &  23.06 &    --  &  4.96 & 0.1768200 &  103. &
   33.\\
 189 & 14.0876 &   ---  &  23.57 &  24.53 &    --  &  4.47 & 0.2951362 &  103. &
   33.\\
 215 & 14.1534 &   ---  &  23.46 &  23.86 &    --  &  2.05 & 0.0709211 &  113. &
   35.\\
 221 & 14.1517 &   ---  &  22.92 &  22.93 &    --  &  6.73 & 0.3010899 &  127. &
   42.\\
 228 & 14.1330 &   ---  &  20.74 &  20.95 &    --  &  4.16 & 0.0275558 &  113. &
   35.\\
 241 & 14.0400 &   ---  &  21.42 &  22.57 &  20.56 &  7.65 & 0.1887201 &  116. &
   38.\\
 285 & 14.1491 &  0.602 &  21.75 &  23.85 &    --  &  6.12 & 0.1252796 &  112. &
   34.\\
 313 & 14.0497 &  0.797 &  21.69 &  22.79 &  20.59 &  6.16 & 0.1268137 &  122. &
   34.\\
 331 & 14.0956 &   ---  &  22.88 &  23.41 &  20.73 &  4.90 & 0.1729592 &  112. &
   37.\\
 334 & 14.1615 &   ---  &  20.89 &  22.04 &  19.64 &  8.91 & 0.1203090 &  101. &
   33.\\
 334 & 14.1602 &   ---  &  21.19 &  21.90 &  20.05 &  6.81 & 0.1527288 &  101. &
   33.\\
 363 & 14.1002 &   ---  &  21.10 &  21.44 &    --  &  7.61 & 0.1869484 &  107. &
   35.\\
 379 & 14.1444 &  0.742 &  22.25 &  23.88 &  20.13 &  4.24 & 0.1325431 &  118. &
   35.\\
 419 & 14.0019 &   ---  &  23.15 &  23.42 &    --  &  2.80 & 0.1282332 &  130. &
   35.\\
 444 & 14.1609 &   ---  &  18.93 &  20.83 &  18.30 &  8.95 & 0.0260458 &  133. &
   35.\\
 454 & 14.0224 &   ---  &  22.57 &  23.58 &    --  &  5.21 & 0.1932082 &  115. &
   34.\\
\hline \hline
\end{tabular}
\end{center}
Notes to Table 1\\
$^*$ Radiosource for which probability has been calculated from radio counts. \\
$^1$ Some of these redshifts are not from the original CFRS catalogues
or were subsequently changed after further analyses: the redshift for
CFRS 14.1157 was reported as z = 1.13 by Hammer et al. (1995), and the
redshift for 14.1042 was given as z = 0.7217 by Lilly et al. (1995b).
There is still some uncertainty in both of these. ``Redshifts'' for the
seven stars (denoted by `` star '') were determined from additional
spectra. `` ---  '' indicates that the redshift is unknown.\\
$^2$ `` -- '' indicates that no observation is available (spectroscopy or K
imagery) \\
$^3$ Distance in arcsec between ISO source position and optical counterpart\\
$^4$ Probability on non-coincidence. \\
$^5$ Flux in $\mu$Jy (aperture of 9\arcsec).\\
}
\end{table*}

\begin{figure}
  \leavevmode
  \psfig{file=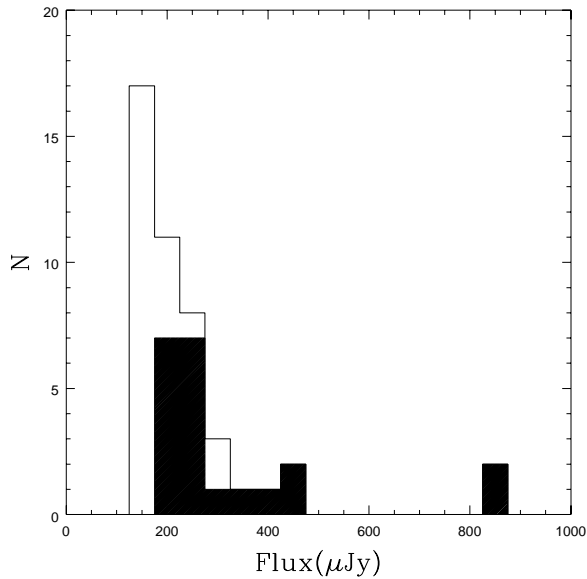,height=8cm,width=8cm}
  \caption{The  5--8.5 micron flux distribution of the
    ISOCAM sources. The ``secure'' identifications listed in
    catalogues 1 and 2 are shown
    in the black shaded area. }
\end{figure}

\section{Catalogue and Identification of Counterparts}
\subsection{Astrometry}

To derive accurate positions of the ISO sources is limited by
(a) the large pixel size (6\arcsec) of the initial image, (b) the size
of the diffraction disk (2 to 3\arcsec), and (c) errors in the data
reduction procedure (including those due to distortion corrections).
The astrometric accuracy of the CFRS galaxies in 1415+52
field is 0\farcs15 relative to radio positions (Hammer et al. 1995).  
The overall centering of the ISOCAM field was based on the positions of the
5 brightest stars and the bright CFRS14.1157 source.  The resulting
differences between the optical and ISO positions for all sources with
counterparts (see next section) are shown in Fig. 4.  All but 9 of the
sources are within a radius equivalent to one pixel (6\arcsec) in the
original ISOCAM data.  From comparison of the positions of optical and
ISOCAM LW2 sources, the median difference is $\sim$4\farcs2.

\begin{figure}
    \psfig{file=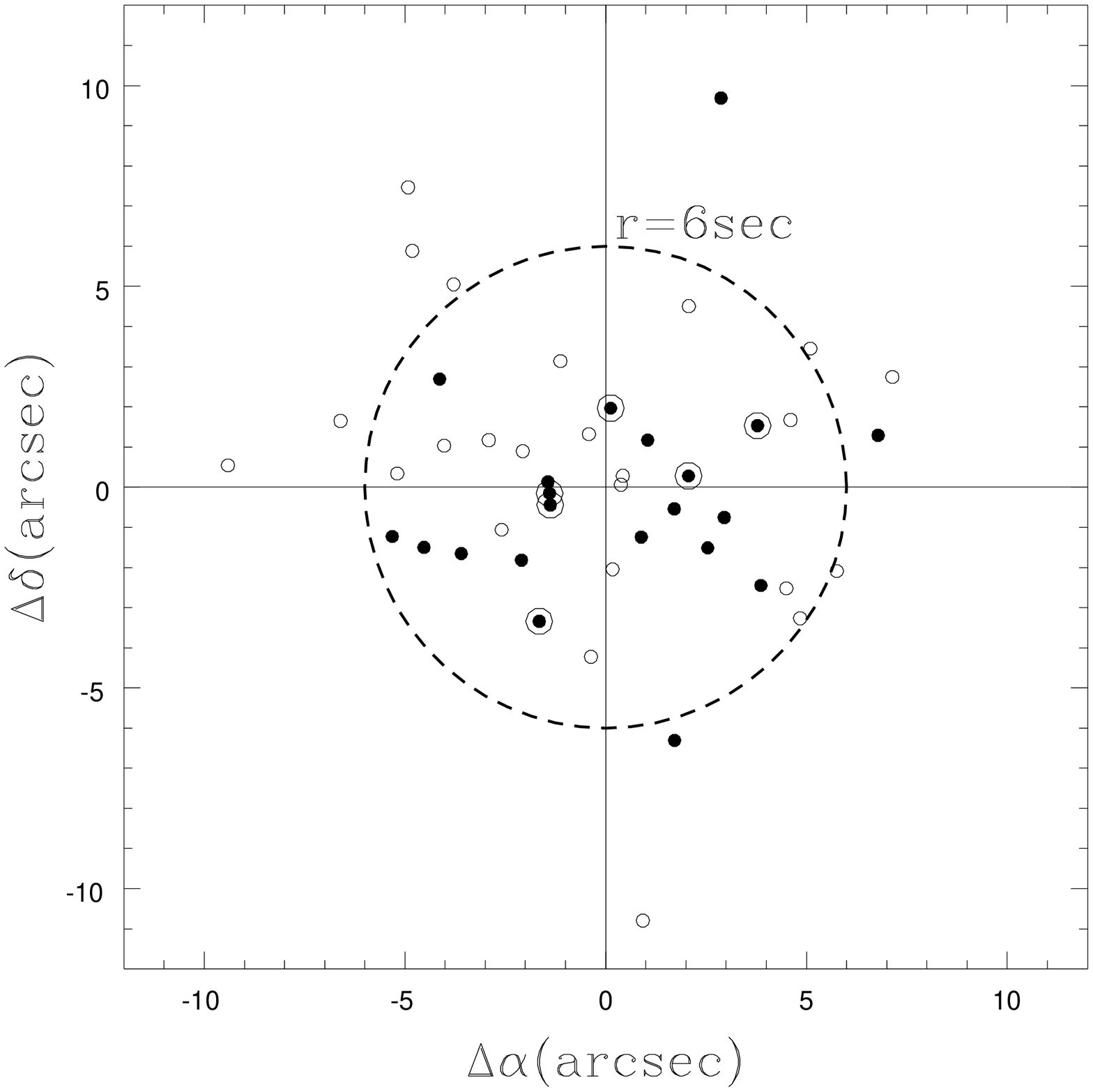,height=6cm,width=6cm}
    \caption{Distances between ISO sources and their optical
      counterparts. The six sources used to determine the relative
      ISO and optical field centers are indicated by circled points.
      Full dots indicate sources with S/N $\geq$ 4, open dots, sources
with 4$>$S/N$\geq$ 3.}
 \end{figure}

\subsection{Optical counterparts}

 We have first compared the 6.75$\mu$m frame with $\mu$Jy radio sources (Fomalont et al, 1991), and have calculated the probability of a pure coincidence, assuming Poisson statistics:

\begin{equation}
P(d, S_{5GHz}) = 1 - e^{-n(S_{5GHz}) \pi d^2}  ,
\end{equation}

where d is the angular distance between the ISOCAM LW3 source and the radio
source in degrees and n is the integrated density of radiosources at flux 
$S_{5GHz}$ ( n$^* (S_{5GHz}$) = 83520 $S_{5GHz}^{-1.18}$).
Six ISOCAM LW2 sources are thus identified (see Table 1) with their optical
counterparts  derived from Hammer et al (1995).

For all the other sources without radio counterpart we have used the $I_{AB}$
band counts. All optical sources within 12\arcsec\ of an ISOCAM LW2
source have been considered.  Assuming Poisson statistics, the
probability density of a pure coincidence between an ISOCAM and optical
source is:

\begin{equation}
 Prob(d, I_{AB}) = 1 - e^{-n(I_{AB}) \pi d^2},
\end{equation}

\begin{figure}
    \psfig{file=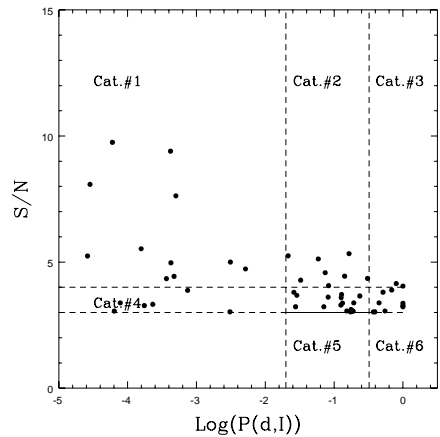,height=6cm,width=6cm}
    \caption{Relationship between the S/N of the ISOCAM detection and
      the probability (P) that the identification is purely by chance. The
      lines show the location of the six
      catalogues described in Tables 1 and 2.}
 \end{figure}

\noindent
where d is the angular distance between the ISO object and the optical
source in degrees and n($I_{AB}$) is the integrated density of galaxies at 
magnitude $< I_{AB}$ from the analysis by Lilly et al. (1995c).  Figure 5 
shows the relationship between the S/N of the ISOCAM observation and the 
probability (P) that the optical source is projected by pure coincidence. 

We are aware that possible non gaussian positional uncertainties
related to extended emission, source confusion and residuals can affect
our probability calculations for low S/N sources. As shown in Fig. 6,
some sources present asymmetric shapes, and for few of them, only a part
of the ISOCAM source overlaps the optical counterpart. Indeed the final
noise structure is far from being gaussian, due to possible glitch residuals, 
 and the ISOCAM position has not been corrected for possible image distortions.
To calibrate our probabilities in an empirical way,  we have applied a
 random match control test to the ISO field by rotating it by
90, 180 and 270 degrees relative to the optical image. Only 2$\pm$1
of the 54  ISOCAM sources are found randomly associated to an optical
counterparts ($I_{AB}<$22.5 and P$<$0.02), which should be compared to the 22
 P$<$0.02 counterparts displayed in Table 1, this strengthens our probability
 calculations. Non gaussian  cannot affect our probability calculations 
by more than a factor 2. 

\subsection{Source catalogs}
As illustrated in Fig. 5, six catalogues were
constructed which contain identifications of varying degrees of
confidence (from 1 -- highest, to 6 -- no identification) according to
their probability (P) and S/N ratio.  Table 1 lists the identifications
of ISO sources in four of the catalogues (in decreasing confidence
level in each successive catalogue).  The ISOCAM LW2 sources and their
associated optical sources are listed in columns 1 \& 2.  Column (3)
gives the redshift if available. In the redshift column , ``star'' indicates 
that the object has been identified spectroscopically
as a star, `` --  '' indicates that no redshift is
available. Columns (4) (5) and (6) give the $I_{AB}$, $V_{AB}$ and
$K_{AB}$ magnitudes (a `` --- '' indicates that no photometry is
available); column (7) lists the angular distance (in arcsec) between
the ISOCAM source and optical counterparts; column (8) gives the
associated probability of coincidence; and columns (9) and (10) give
the flux at 5--8.5 $\mu$m and the error (in $\mu$Jy).

The 21 sources listed in catalogues 1 and 2 are considered to have
secure identifications since they are relatively strong sources (S/N
$>$ 4) and have a relatively low probability of chance coincidence.  All but
two of the counterparts have $I_{AB} < 22.5$, the limit of the CFRS
spectroscopic survey, and  spectra are available for 13 of these
counterparts.

The 24 sources in the supplementary catalogues 4 and 5 are
identifications with lower S/N sources (4$\geq$S/N$\geq$ 3).  In five
cases in the ``least secure'' catalogue 5,  more than one optical
counterpart is listed as possibly being associated with an ISOCAM
source.  In most cases, the more probable counterpart is optically
brighter than, or as bright as, the alternate, providing additional
evidence that it is the likely counterpart.  In the following analysis
we use only  the ``best'' optical identifications (i.e., those with the
lowest probability of chance coincidence), but  all counterparts with a
probability within a factor two of the smallest probability are
retained in Table 1 for reference.  Statistically, 1.5 out of the 18
sources in catalogue 5 are expected to be purely chance coincidences.
As well as having fainter mid-IR fluxes, the optical counterparts of
these 24 sources are also fainter optically on average (8 are fainter
than $I_{AB} = 22.5$).

Figure 6 shows the superposition of  ISOCAM LW2 source contours and the
optical counterparts  of 14 extragalactic sources (from catalogues 1,
2, 4 \& 5) with known redshifts.  The optical image was extracted from
a composite $B+V+I$ image centered at the ISO position.

\begin{figure}
  \begin{center}
    \leavevmode
     \psfig{file=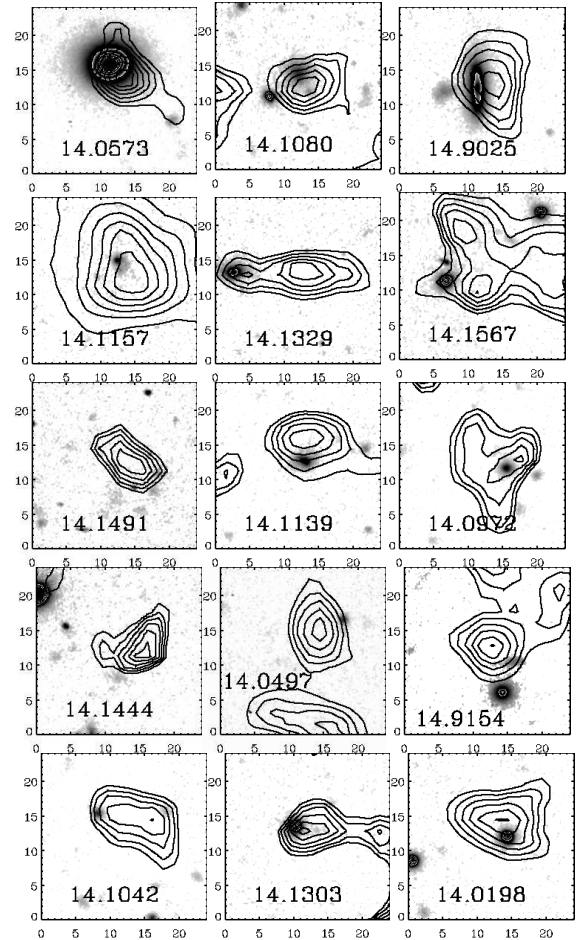,height=12.5cm,width=7.5cm}
   \end{center}
  \caption{Charts of images  (24\arcsec $\times$24\arcsec) centered
    on the ISOCAM LW2 sources with known redshifts from catalogues 
    1,2, 4 and 5.  ISOCAM LW2 contours $>3 \sigma$  are overlaid.
Glitch residuals combined with the large ISOCAM pixel size, can create 
 apparently (and artificially) extended sources. This complicates 
in some cases, the identification of the optical counterpart.}
\end{figure}

\begin{table}
  \begin{center}
    \caption{ISOCAM LW2 sources without optical counterparts. Those sources above the line are in catalogue 3 (S/N~$>$~4) and those below comprise catalogue 6 (4~$>$~S/N~$>$~3).}
    \leavevmode
    { \scriptsize
    \begin{tabular}{crcrrcrcc} \hline
ISO  &  \multicolumn{3}{|c|}{$ \alpha_{2000}$} & \multicolumn{3}{|c|}{$\delta_{20
00}$} & Flux &Error\\  \hline \hline
096  &14 &18 & 4.2 &52 &26 &27.7 &  166 & 41 \\
375  &14 &17 &51.3 &52 &31 & 9.9 &  141 & 34 \\ \hline
113  &14 &18 & 6.1 &52 &28 & 8.4 &  137 & 36 \\
119  &14 &17 &41.7 &52 &34 &56.4 &  128 & 38 \\
195  &14 &17 &28.4 &52 &29 &55.5 &  101 & 33 \\
257  &14 &17 &33.7 &52 &33 &22.2 &  113 & 35 \\
333  &14 &18 &16.3 &52 &34 &55.0 &  108 & 33 \\
369  &14 &18 &19.0 &52 &31 &33.7 &  115 & 34 \\
380  &14 &18 & 6.7 &52 &25 &29.7 &  183 & 47 \\  \hline \hline
    \end{tabular}
    }
  \end{center}
\end{table}

Among the 54 ISOCAM LW2 sources with S/N $\ge$ 3, 9 were not identified
to the limit of our combined $B+V+I$ image. Unfortunately, we have $K$
band images for only a small portion of the field, so it is not
possible to determine whether all sources are visible in the
near-IR and what their colors are.  The positions and fluxes of the 9
``non-identified'' ISO sources are listed in Table 2.

In summary, 45 of the 54 ISOCAM LW2 sources (83\%) were identified with
galaxies and stars on composite $B+V+I$ images ($I_{AB} \loa 23.5$) of
the CFRS1415+52 field. 21 of these have secure identifications with
optical counterparts (including 7 stars). As shown below, some of the
visible-light properties of the 24 less secure counterparts add
considerable support to their identifications.  In addition to the 7
spectroscopically confirmed stars, three other sources  have profiles
of compact sources (stars, QSO's, etc) on our $I$ band image, so that
10 of the 45 may be stars (although the three could also be QSOs, from
B,V and I band images and colors).

\section{Optical properties of ISO counterparts}
\subsection{ Photometry}

The photometric properties of the identified optical counterparts are
illustrated in Figs. 7, 8 and 9. Figure 7 shows that the less secure
counterparts are fainter in $I_{AB}$ on average than the secure counterparts.
In addition to the 17\% of ISOCAM LW2 sources that have no obvious
optical counterparts, the counterparts for another 18\% are fainter
than the limit (I$_{AB}$ = 22.5) of CFRS spectroscopy, and hence some
of these may be at z$>$1. Figure 8 compares  the $(V-I)_{AB}$
distribution of all the secure and less secure extragalactic
identifications (lower panel) with the same distribution for all CFRS
galaxies ($17.5 \leq I_{AB} \leq 22.5$) (upper panel).  This diagram
shows that  ISOCAM LW2 and CFRS sources have the same median
$(V-I)_{AB}\sim$ 1), which corresponds to the color of an Sbc in the
redshift range from z = 0.3 to 1.  Figure 9 shows that the median of the
$(I-K)_{AB}$ histogram of the ISOCAM sources identified with known
galaxies is 0.5 mag redder than that of CFRS galaxies, which is
$(I-K)_{AB} \sim 1.3$ mag.  Five of the sources (114, 331, 379, 461,
477) are ultra-red  (with $(I-K)_{johnson} \geq 3$). One of these is at
z = 0.7, but the other four are possible high redshift objects.

\begin{figure}
  \begin{center}
    \leavevmode
    \psfig{file=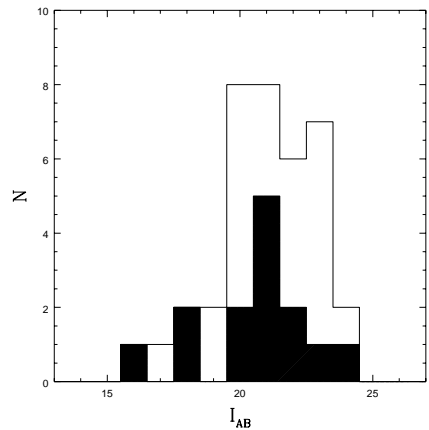,height=6cm,width=6cm}
  \end{center}
  \caption{The distribution of $I_{AB}$ for all ISOCAM LW2 sources
    with optical counterparts. The
    black shaded area denotes secure identifications from catalogues 1 and 2.}
\end{figure}

\begin{figure}
  \begin{center}
    \leavevmode
    \psfig{file=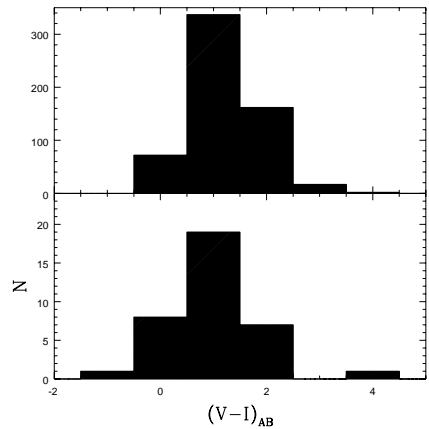,height=6cm,width=6cm}
  \end{center}
  \caption{The distribution of $(V-I)_{AB}$ for all ISO
    sources identified with extragalactic sources compared with that for
    all CFRS galaxies (top panel)}
\end{figure}

\begin{figure}
  \begin{center}
    \leavevmode
    \psfig{file=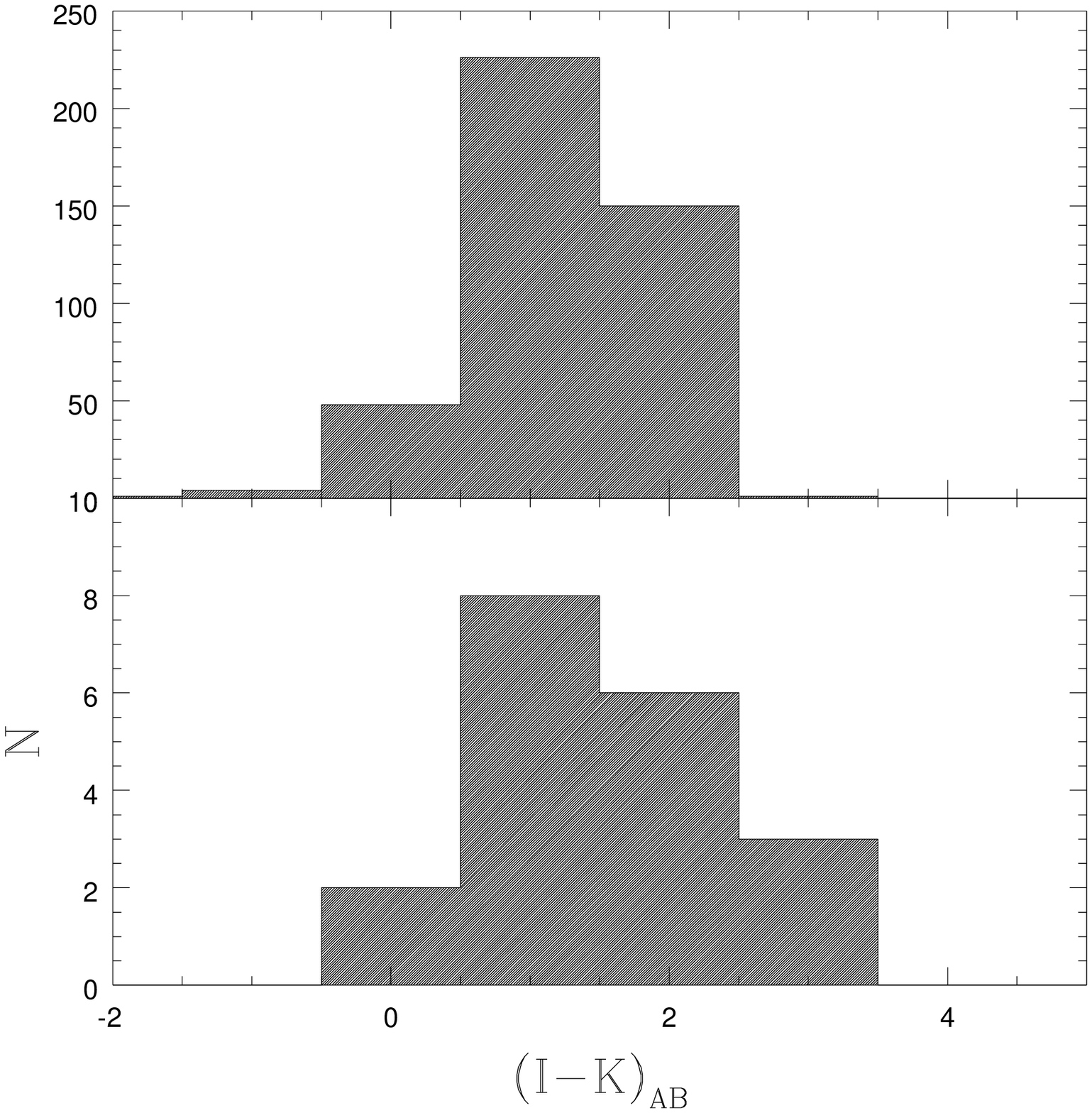,height=6cm,width=6cm}
  \end{center}
  \caption{The distribution of
    $(I-K)_{AB}$ for sources identified with extragalactic
    objects compared with that for all CFRS galaxies (top panel).}
\end{figure}

\subsection{Spectra}

Spectra for 22 of the 45 optical counterparts of our ISOCAM LW2 sources
are available in the CFRS database (including 7 stars, generally M or K
stars). Several methods were used to classify the objects through their
spectra and other properties:  diagnostic diagrams (using
O$_{II}$/H$_{\beta}$, O$_{III}$/H$_{\beta}$, H$_{\beta}$,
S$_{II}$/H$_{\alpha}$, N$_{II}$/H$_{\alpha}$ ratios), radio properties
(Hammer et al. 1995), star formation rate (O$_{II}$3727\AA~equivalent
widths) and the Balmer index D(3550 $-$ 3850) defined as
$f_{\nu}$(3750-3950)/$f_{\nu}$(3450-3650) at rest (see
Hammer et al. 1997) The latter is an indicator of recent star formation,
and is very well correlated to the H$_{\delta}$ equivalent width. Large
values of D(3550 $-$ 3850) ($>$ 0.18) indicate significant contribution of A star population (such objects are classified as '$S+A$'), suggesting a major burst of star formation some 0.5 Gyr ago. Altogether, we find: eight $S+A$ galaxies (Balmer index, D(3550 $-$ 3850) $>$ 0.18 and/or detection of W$(H_{\delta}) \leq -5$\AA, absorption), one starburst (diagnostic diagram), four QSO and Seyfert 1 galaxies (from their continuum flux and broad lines) and two Seyfert 2 galaxies (based on the diagnostic diagrams).

\begin{figure*}
    \psfig{file=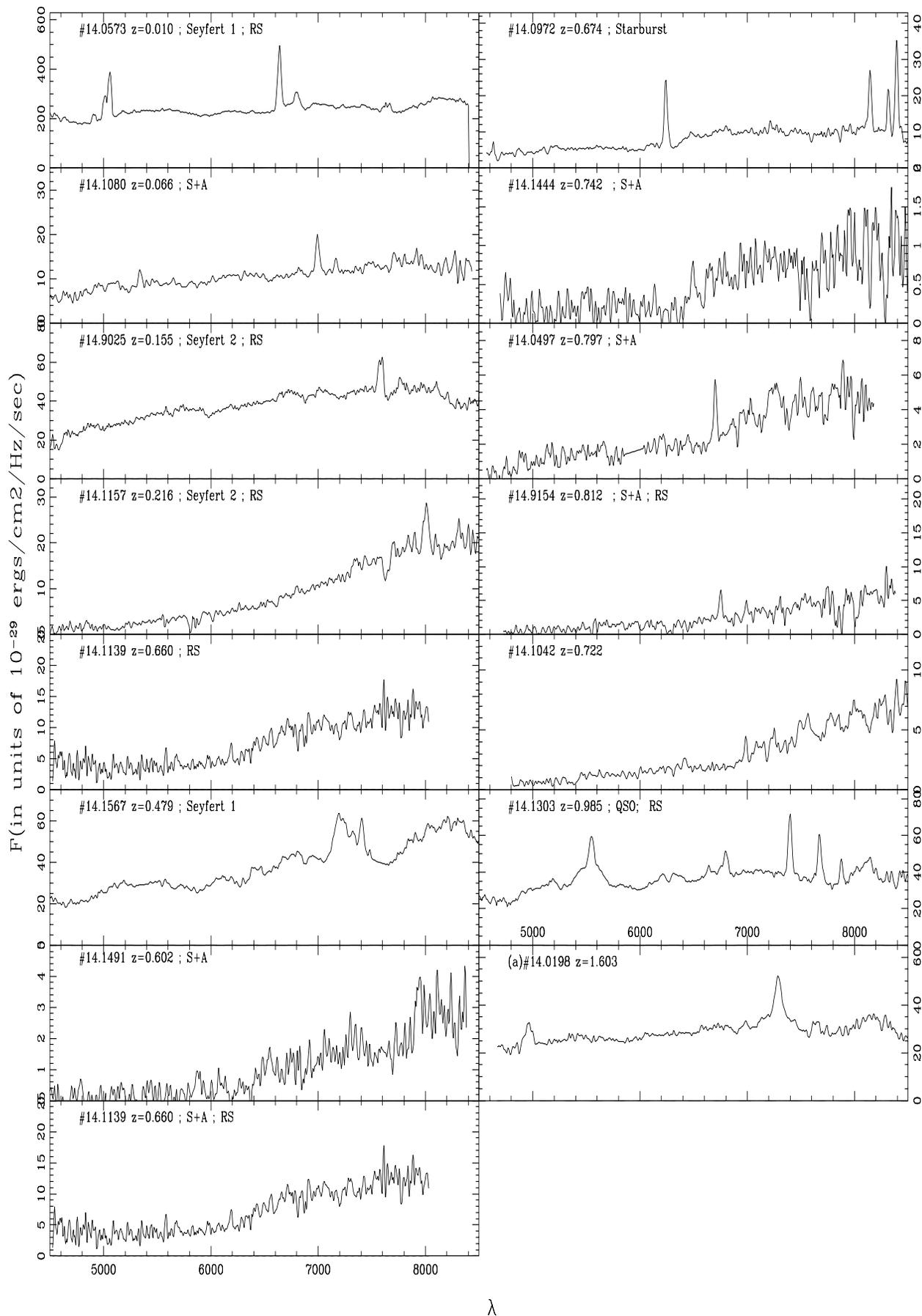,height=25cm,width=17cm}
    \caption{Spectral classification of the galaxy spectra from their
      continuum and line properties. From top left, the first four are
      QSO and Seyfert 1, the next two are Seyfert 2, the next 8 are
      classified as $S+A$, and the last one (bottom right) as a
      starburst galaxy.}
\end{figure*}

\begin{table}
\caption{ISOCAM LW2 galaxies with spectra.}
\begin{center}
\leavevmode
\begin{tabular}{lcccccl}
CFRS    & z     & C$^{1}$ &  M$_B$ & W$_{OII}^{2}$& S$^{3}$&Type\\
\hline \hline
14.0573 & 0.010 & 4  & $-$16.79 &   --    & 19 & Seyfert 1 \\
14.1080 & 0.066 & 4  & $-$17.17 &   --    & -- &  $S+A$ \\
14.9025 & 0.155 & 1  & $-$20.34 &   --    & 30 & Seyfert 2 \\
14.1157 & 0.220 & 1  & $-$18.55 &   --    & 77 & Seyfert 2  \\
14.1329 & 0.375 & 5  & $-$21.41 &  15.0   & 171& $S+A$ \\
14.1567 & 0.479 & 1  & $-$22.31 &   --    & -- &  Seyfert 1 \\
14.1491 & 0.602 & 5  & $-$20.25 &  59.0   & -- & $S+A$ \\
14.1139 & 0.660 & 4  & $-$22.27 &  19.0   & 141& $S+A$ \\
14.0972 & 0.674 & 1  & $-$21.50 &  40.0   & -- & Starburst  \\
14.1444 & 0.742 & 5  & $-$20.47 &  17.0   & -- & $S+A$  \\
14.0497 & 0.797 & 5  & $-$21.28 &  29.0   & -- & $S+A$ \\
14.9154 & 0.812 & 1  & $-$21.41 &  39.0   &  55& $S+A$ \\
14.1042 & 0.875 & 2  & $-$21.11 &  27.0   & -- & $S+A$ \\
14.1303 & 0.985 & 4  & $-$23.48 &  --     & 78 & QSO \\
14.0198 & 1.603 & 1  & $-$24.72 &  --     & -- & QSO \\
\hline \hline
\end{tabular}
\begin{tabular}{p{0.4\textwidth}}
Notes to Table 3:\\
(1) ``Confidence catalogue'' number. \\
(2) Rest frame equivalent width of OII in \AA.\\
(3) S$_{1.4GHz}$ flux in $\mu$Jy.
\end{tabular}
  \end{center}
\end{table}

The spectra, divided according to the above classifications, are
shown in Fig. 10 and relevant data for these sources are listed in Table 3.
The optical counterparts of the eight less secure (catalogues 4 \& 5)
ISOCAM sources include a large fraction (6 of 8) of $S+A$ galaxies. Our confidence in the reliability of the identifications is strengthened by
 the fact that five of them are also detected either at radio wavelength
or at 15$\mu$m with S/N $\geq$ 4 (Flores et al. 1998, submitted).

\begin{figure}
    \psfig{file=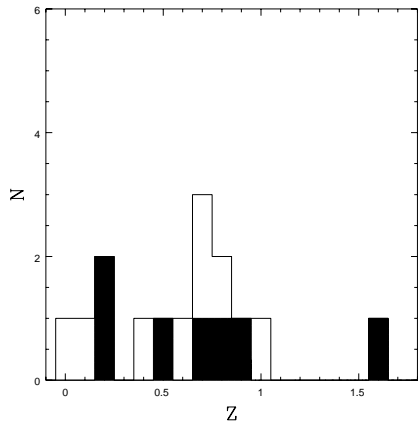,height=6cm,width=6cm}
    \caption{The redshift histogram of the 15 identifications with spectra from
      the CFRS database. The black shaded area represents objects from
     catalogues 1 and 2. The two highest redshift objects are QSOs.}
\end{figure}

As shown in Figure 11, the median redshift of the 15 extragalactic
sources is z $\sim$ 0.60, close to the average CFRS value (this is not too surprising since
our spectroscopic sample of ISOCAM sources has the same optical flux
limit as the CFRS).  The two most distant objects are  QSOs at z $\sim$
1 and 1.6 (Schade et al. 1995).  The latter has the highest redshift in
this CFRS field.  Most of the
strongest sources with secure identifications (catalogues 1 \& 2)
are AGNs (QSO and Seyfert).

\section{Summary of ISO identifications}

A total of 54 mid-IR sources were detected in a deep survey of the CFRS
1415+52 field with ISOCAM.  The reliability of the detections of these
predominantly very faint sources has been strengthened through
comparison of the results of two completely independent analyses of the
data.  The counts at 6.75$\mu$m are in a very good  agreement with
those of the Deep ISOCAM Survey (Lockman-Hole Survey) which is $\sim$
0.22 counts per sq. arcmin for $S_{6.75\mu m} > 250\mu$Jy. Further support for
the reliability of the identifications is provided by the properties of
their optical counterparts:
\begin{itemize}
\item In the CFRS 1415+52 field, there are only 25 radio-sources with
 $S_{5GHz} > 16\mu Jy$ (Hammer et al. 1995)  and 34 ISOCAM sources with
 $S_{6.75\mu m} > 150\mu Jy$ which are identified with I$_{AB}
 <$22.5 galaxies (out of a total of $\sim$600 I$_{AB}<$~22.5 objects). 
 Based only on optical--IR astrometry six ISOCAM sources are associated with
 radiosources, and the cumulative probability of
 finding 6 or more radiosources in this sample by chance is only 0.0037.
\item Among the 15 galaxies with spectra, 6 (40\%) are found to be AGNs,
even though only 7\% of CFRS galaxies are AGN (Hammer et al. 1997).
\item Of the star--forming galaxies, 7 out of 8 have a Balmer index 
larger than the median value of CFRS galaxies, possibly implying a
relationship between the presence of dust and the A-star population.
Also dusty galaxies present larger color indices (Flores et al., 1998)
\item The ISO galaxies  have a median $(I-K)_{AB}$ color
 which is 0.5 mag redder than randomly-selected CFRS galaxies and 5 of
 them have a very red $(I-K)$ color. 
\end{itemize}
Clearly, the ISOCAM LW2 sources have different average properties than  a 
random sample of CFRS galaxies, giving additional
confidence in our detections, even down to $150\mu Jy$.

\section{Discussion: the nature of the 6.75$\mu$m sources}

A total of 54 $S_{6.75\mu m} > 150\mu Jy$ were detected in the
10\arcmin~$\times$~10\arcmin\ field. Of these, 9 could not be
identified with any optical counterparts to the limit of our optical
images ($I_{AB} \sim 23.5$); two sources with S/N $\sim$4 and seven
more with S/N $>$ 3 remain unidentified. Of the remaining 45, there are
21 ``secure identifications'' (catalogues 1 \& 2). Optical spectra are
available for 14 of these showing that 7 are stars (mostly K and M) and
5 out of the 6 remaining sources display AGN or starburst activity.
Thus, most of the strongest extragalactic sources with secure
identifications are AGNs.  The  mid-IR flux in AGN host galaxies is
believed to come directly or indirectly from the hot dust in the
inner regions of a torus around the active nucleus.
 Two of the ``secure identifications'' have $I_{AB} > 22.5$,
the limit of the CFRS spectroscopic survey.

Analysis of the properties (photometric, spectroscopic, radio) for all
sources with optical counterparts indicates that a large fraction of
even the ``less secure'' counterparts are likely to be the correct
identifications. There are 45 identified sources in total and 7 are
spectroscopically confirmed stars, leaving 38 possible extragalactic
sources (although 3 of these have stellar profiles). Ten of these have
$I_{AB} > 22.5$ and are fainter than the CFRS survey limit, but spectra
are available for 14 of the remaining 28 counterparts. As Table 3
indicates, 53\% are classified as $S+A$, showing evidence of an A star
population and star formation activity, indicative of significant star
formation $\sim$ 0.5 Gyr ago.  Detection of $S+A$ galaxies in the
mid-IR is supported by preliminary results from ISOCAM LW3 15$\mu$m
data (Flores et al. 1998). 

The average ratio between the energy ($\nu F_\nu$) at $6.75/(1+z)$ $\mu$m and the visible energy (0.835/(1+z) $\mu$m) is high, 2.1 $\pm$ 0.7 in average,  for the 9 galaxies exhibiting star formation activity.  
It is higher that the value estimated for a local starburst galaxy
after reshifting it to the median redshift (z=0.71) for our sample of
star forming galaxies, for which we find $\nu F_\nu(LW2)/\nu
F_\nu(I_{AB}$=0.63. Figure 12 shows the redshift distribution of this
ratio, compared to local templates  (starburst, elliptical and QSOs from Schmitt et al.  1997). Although several objects appears compatible with local galaxies, Fig. 6
indicates that Mid-IR to optical flux ratio is not sufficient to
determine the nature of the source emission. On the other hand, some sources appear redder than any kind of local templates. This does not seem
related to a deficiency in our flux calibration at 6.75$\mu$m (stars do
not show the same excess when compared to Rayleigh-Jeans emission distribution), and the excess is apparently higher than our expectations for photometric errors. Indeed there are several bright ISOCAM sources (including the brightest one, CFRS14.1157) which present this red excess. From a purely statistical point of view, this is not unexpected,
since the ISOCAM detections correspond to the small fraction (5\% ) 
of I$_{AB} <$ 22.5 galaxies at $0 < z < 1.6$ which are the most extreme mid-IR emitters. On the other hand this might indicate that several field galaxies up to z=1 have higher Mid-IR fluxes (related to PAH or to hot dust) when  compared to local galaxies.\\

\begin{figure}
    \psfig{file=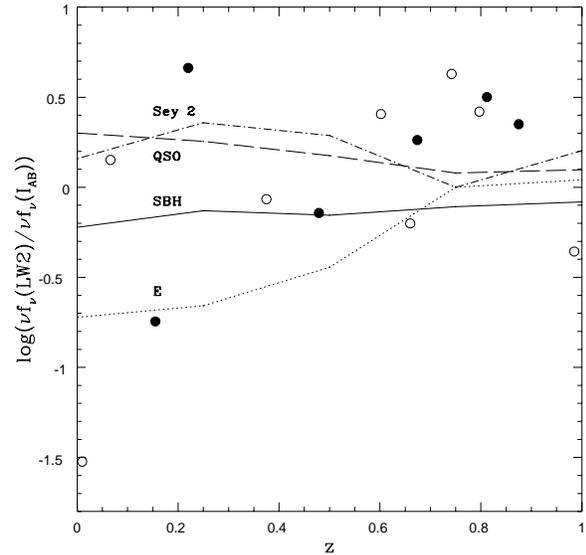,height=8cm,width=8cm}
    \caption{Redshift distribution of the $\nu F_\nu$ ratio for 14 extragalactic
sources. Full dots represent secure ISOCAM sources (S/N $>$ 4, open dots those
which are less secure. Are also displayed curves derived from
redshifted local templates (from Schmitt et al.  1998; SBH:
 reddened starburst).}
 \end{figure}

Some of the unidentified sources and some of the sources with
counterparts for which we do not yet have spectra may be at z $>$ 1.
The fact that a) many of the counterparts without spectra are fainter
than average and b) that some of these are very red in $(I-K)$ supports
this possibility. On the other hand, they may also simply be highly
reddened. A crude estimate of the number of high z ($z>1$) mid-IR
emitters can be made by adding the following:
\begin{itemize}
\item The one spectroscopically confirmed I$_{AB} \leq$ 22.5 
  extragalactic ISO source at $z>1$.
\item The three (30\%) of the ten 22.5 $<$ I$_{AB} <$ 23.5 ISO sources that are likely to be at  $z>1$ (from the extrapolation by Lilly et al. 1995c)
\item All nine unidentified sources.
\end{itemize}
In other words, the maximum number of ISO sources which could be at
z $>$ 1 is 14, or 30\% of the 44 $S_{6.75\mu m} > 150\mu Jy$ sources which are non-stellar (7 "spectroscopic" stars and 3 "stellar profile" objects excluded).

Perhaps the most intriguing object in our ISO sample is CFRS14.1157,
which is the brightest non-stellar object in our LW2 image. It is
likely to be a heavily absorbed AGN at z = 0.216.  Further observations
of this source are warranted. Deeper and more complete near-IR imaging
and spectroscopy of a number of the optical counterparts and the
unidentified sources would also be of interest.

\noindent
{\it Acknowledgments:}
The ISOCAM data presented in this paper were analyzed using software
developed by F.X. D\'esert  and ``CIA'' (a joint development by the
ESA Astrophysics Division and the ISOCAM
Consortium led by the ISOCAM P.I., C. C\'esarsky, Direction des Sciences de la Matiere, C.E.A., France). We thank  D. Pelat
for useful discussions about the identification procedure. We are grateful to the referee for suggestions and comments which help us to seriously improve the manuscript.

\end{document}